\documentstyle[11pt,equation,epsf]{article}
\input{psfig.sty}
\begin{document}
\newcommand{\mpl}{M_{Pl}}
\newcommand{\lx}{\lambda}
\newcommand{\ex}{\epsilon}
\newcommand{\be}{\begin{equation}}
\newcommand{\ee}{\end{equation}}
\newcommand{\een}{\end{subequations}}
\newcommand{\ben}{\begin{subequations}}
\newcommand{\beq}{\begin{eqalignno}}
\newcommand{\eeq}{\end{eqalignno}}
\def \lta {\mathrel{\vcenter
     {\hbox{$<$}\nointerlineskip\hbox{$\sim$}}}}
\def \gta {\mathrel{\vcenter
     {\hbox{$>$}\nointerlineskip\hbox{$\sim$}}}}
\pagestyle{empty}
\noindent
\begin{flushright}
CERN-TH/97-163 \\
astro-ph/9707214
\end{flushright} 
\vspace{3cm}
\begin{center}
{ \Large \bf
Fine Tuning of the Initial Conditions \\
\vspace{0.2cm}
for Hybrid Inflation
} 
\\ \vspace{1cm}
N. Tetradis \\
\vspace{1cm}
CERN, Theory Division, \\ 
CH-1211, Geneva 23, Switzerland \\
\vspace{2cm}
\abstract{
We study the evolution of 
regions of space with various initial field values 
for a simple theory that can support 
hybrid inflation. Only very narrow domains 
within the range of initial field values below the Planck scale
lead to the onset of inflation. 
This implies a severe fine tuning for the initial configuration 
that will produce inflation.
\\
\vspace{1cm}
PACS number: 98.80.Cq
}
\end{center}
\vspace{6cm}
\noindent
CERN-TH/97-163 \\
July 1997

\newpage
\setlength{\baselineskip}{20pt}

\pagestyle{plain}
\setcounter{page}{1}

One of the most attractive feature of hybrid inflation 
\cite{hybrid} is that 
it can take place for values of the inflaton field below the 
Planck scale \cite{cop}. This permits the construction of realistic 
field theoretical models without the need of a description of 
quantum gravity. Several such models have been proposed within the
context of supersymmetry \cite{cop}-\cite{gia}. 
Inflation
requires the presence of an almost flat direction in field space
with non-zero vacuum energy. A very small slope along this direction
results in the slow rolling of the inflaton field $\phi$. For hybrid
inflation, the slow rolling 
occurs along a valley of the
effective potential. It ends when the valley turns into a ridge and 
the slow-roll regime terminates, owing to the growth of
fluctuations of fields orthogonal to 
the inflaton field. Typically this point of instability
corresponds to a value $\phi_{ins}$ of the inflaton below the 
Planck scale. 

The question we would like to address is whether generic initial
values of the fields below the Planck scale can 
lead to the onset of inflation.
Clearly, there is an area where this is likely, the almost
flat direction. Our aim is to explore the whole field space 
below the Planck scale and
establish whether inflation is a frequent occurrence or a rare
event. Similar investigations have already been performed in refs.
\cite{first1,first2}. Our main observation is that the 
field range that can lead to inflation is much more constrained
than what one would naively
infer from refs. \cite{first1,first2}, unless a significant
fine tuning of the fields is imposed within the space regions that
eventually inflate.

The onset of inflation requires a region of
space, the size of a few Hubble lengths,
where the fields take almost constant values, so that the
gradient energy density is negligible compared to the potential energy 
density. 
The earliest time at which one could start talking about such 
regions of space 
is when the Planck era (during which
quantum gravitational fluctuations dominate) ends and 
classical general relativity starts becoming 
applicable. The initial field values 
within each region
are expected to be of order $\mpl$, 
or larger if the couplings
of the theory are small, as in chaotic inflation \cite{chaotic}.
In this work we concentrate on field values below $\mpl$ for two reasons:
a) Hybrid inflation does not require extremely small couplings.
b) In general, the field-theoretical models for which a potential can be
reliably computed
assume an expansion in powers of $\phi/\mpl$.  

Inflation could start at the end of the Planck era if
the fields take values in the parts of the potential that are able to 
support it. However, there is also the possibility that the 
fields will evolve, from some initial values that do not give
inflation, to different values that do. We consider all possible field values, 
assuming an initial degree of homogeneity over a few Hubble lengths,
and follow the evolution by integrating the classical equations of motion.
In this way we determine the range of field values that eventually
lead to inflation. We show that, for many realistic models, 
the initial fields must
be extremely homogeneous if their initial values are below the Planck scale.
For example, for the prototype model of 
hybrid inflation (see below), one of the fields 
must be homogeneous with 
at least $\sim 10^{-6}$--$10^{-5}$ accuracy. 
It is very difficult
to calculate a probability distribution for the initial 
field configurations, and thus have a 
quantitative estimate of the probability of inflation to occur. 
This requires a reliable description of Planck scale dynamics. 
For this reason we do not address the question of whether such a homogeneous
initial state is probable or not.
Our results have a much looser interpretation, such as indicating how
``natural'' hybrid inflation is. 

We consider the simplest potential that can support hybrid inflation  
\cite{hybrid}
\be
V(\phi,\sigma)= 
\frac{1}{4} \lx \left(\sigma^2-M^2 \right)^2
+ \frac{1}{2} m^2 \phi^2
+ \frac{1}{2} g \phi^2 \sigma^2,
\label{one}
\ee
with positive mass terms and couplings.
It possesses the symmetries $\phi \rightarrow -\phi$ and
$\sigma \rightarrow -\sigma$. A renormalizable term
$\lx' \phi^4$ is permitted by these symmetries and
should naturally appear. However, 
neglecting such a term can be justified if 
the hybrid inflationary scenario is embedded
in the context of supersymmetry and an appropriate 
superpotential is chosen. 
The term 
$m^2 \phi^2/2$ is also absent in supersymmetric models,  
and the small slope along the
inflationary trajectory is generated by 
non-renormalizable (such as logarithmic)
terms that are 
present when quantum corrections
are taken into account \cite{shafi,giwrgos,gia}.

We use the simplest choice of eq. (\ref{one}) 
for our study. 
Our results do not depend on whether the slope along the inflationary
trajectory is generated by the term 
$m^2 \phi^2/2$ or radiative corrections.
In the concluding paragraphs we discuss a simple superpotential
to which our results apply directly.
We point out, however, that our analysis must
be repeated for the determination 
of the appropriate initial conditions
in supersymmetric inflationary models with 
more complicated superpotentials.

The potential of eq. (\ref{one}) has a valley of minima 
along the $\phi$ axis ($\sigma=0$) for 
fixed $\phi^2 > \phi^2_{ins}$, where 
\be
\phi_{ins} = \pm \sqrt{\frac{\lx}{g}} M.
\label{two} \ee
There is a series of maxima along the 
$\phi$ axis ($\sigma=0$) for 
fixed $\phi^2 < \phi^2_{ins}$. The absolute minima of the
potential are located at ($\phi=0$, $\sigma=\pm M$). 
We assume that a Robertson-Walker metric is a good approximation 
for the regions of space with uniform fields that we are considering. 
The evolution of the fields is given by the standard equations
\beq
H^2 = \left( \frac{\dot{R}}{R} \right)^2 = &\frac{8 \pi}{3 \mpl^2} \left[
\frac{1}{2} \dot{\phi}^2
+ \frac{1}{2} \dot{\sigma}^2 + V(\phi,\sigma) \right]
\label{three} \\
\ddot{\phi} + 3 H \dot{\phi} = & 
- \frac{\partial V (\phi,\sigma)}{\partial\phi} =
-m^2 \phi - g \sigma^2 \phi
\label{four} \\
\ddot{\sigma} + 3 H \dot{\sigma} = & 
- \frac{\partial V (\phi,\sigma)}{\partial\sigma} =
\lx M^2 \sigma  - \lx \sigma^3 - g \phi^2 \sigma,
\label{five} \eeq
with $\mpl = 1.2 \times 10^{19}$ GeV.

Inflation occurs along the $\phi$ axis, where the 
potential is 
\be
V(\phi)= 
\frac{1}{4} \lx M^4
+ \frac{1}{2} m^2 \phi^2.
\label{six}
\ee
In the limit where the second term in the r.h.s. of the above equation
dominates, we obtain
the prototype model of chaotic inflation. This
takes place for values of the field $\phi^2 \gg \mpl^2$.
The COBE observation of the cosmic microwave background
anisotropy leads to the constraint $m \simeq 10^{-6} \mpl$ \cite{chaotic}. 
As we have already explained, we are interested 
in values of the field $\phi^2 \lta \mpl^2$,
for which the first term in the r.h.s. of
eq. (\ref{six}) dominates during inflation. 
This occurs for 
\be
2 m^2 \mpl^2 \ll \lx M^4.
\label{seven} \ee
The ``slow-roll'' parameters \cite{report} are given by \cite{cop}
\beq
\eta(\phi) = &\frac{\mpl^2}{8 \pi}
\frac{d^2 V(\phi)/d\phi^2}{V(\phi)} 
= \frac{m^2 \mpl^2}{2 \pi(\lx M^4 + 2 m^2 \phi^2)}
\simeq \frac{1}{2 \pi} 
\frac{m^2 \mpl^2}{\lx M^4} 
\label{eight} \\
\ex(\phi) = &\frac{\mpl^2}{16 \pi}
\left( \frac{d V(\phi)/d\phi}{V(\phi)} \right)^2
= \frac{\mpl^2 m^4 \phi^2}{\pi(\lx M^4+ 2m^2 \phi^2)^2} 
= 4 \pi \eta^2 \frac{\phi^2}{\mpl^2}. 
\label{nine} \eeq
In the interval $\phi^2_{ins} \leq \phi^2 \lta \mpl^2$, the conditions
for inflation $\ex,\eta \ll 1$
are automatically 
satisfied for the parameter range of eq. (\ref{seven}).

The number of $e$-foldings that occur between the moment that
cosmologically  interesting scales (of the 
order of the present Hubble distance) leave the 
horizon and the end of inflation is $N \simeq 60$. 
We denote by $\phi_{60}$ the value of the inflaton field at horizon crossing
for these scales.
In our model, inflation terminates when $\phi$ rolls beyond the instability
point $\phi_{ins}$. 
In the parameter range of eq. (\ref{seven}),
the relation between $\phi_{60}$ and $\phi_{ins}$ is \cite{cop}
\be
\phi^2_{60} \simeq \phi^2_{ins} e^{120 \eta},
\label{ten} \ee
with $\eta$ given by eq. (\ref{eight}).
The inflationary prediction for the 
spectrum of adiabatic density perturbations at cosmological scales 
\cite{denpert,report} is
\be
\delta^2_H = \frac{32}{75} \frac{V_{60}}{\mpl^4} \frac{1}{\ex_{60}} \simeq
\frac{8 \pi}{75} \frac{\lx^3 M^{12}}{\mpl^6 m^4 \phi^2_{60}}.
\label{eleven} \ee
Comparison with the value $\delta_H = 1.94 \times 10^{-5}$,
deduced from 
the COBE observation of the cosmic microwave background
anisotropy, 
leads to the  constraint \cite{cop}
\be
\frac{M}{\mpl} \simeq 0.13~e^{12 \eta}~g^{-{1}/{10}} \lx^{-{1}/{5}}
\left( \frac{m}{\mpl} \right)^{2/5}.
\label{twelve} \ee
For couplings that are not many orders of magnitude smaller than 1
and a sufficiently small ``slow-roll'' parameter
$\eta$, we obtain a simple relation between
the scales $M$ and $m$
\be
\frac{M}{\mpl} \simeq 0.13
\left( \frac{m}{\mpl} \right)^{2/5}.
\label{thirteen} \ee
The above relation holds as long as our assumption that the first
term in the r.h.s. of eq. (\ref{six}) dominates is valid. 
As we have already mentioned, when the second term 
dominates, this model reduces to the prototype model of chaotic 
inflation, for which the COBE observations imply
$m=10^{-6} \mpl$. 
We expect, therefore, that eq. (\ref{thirteen}) holds as long
as $m \lta 10^{-6} \mpl$. This leads to the constraint
$M \lta 5 \times 10^{-4} \mpl$. A detailed analysis \cite{cop}
verifies this conclusion and predicts a more accurate
bound
$M \lta 2 \times 10^{-3} \mpl$. 
A similar bound applies to all models of hybrid inflation with
couplings of order 1. 
On general grounds one expects the inflationary energy scale
$V^{1/4}$ to be at least two or three orders of magnitude smaller
than the Planck scale \cite{lyth}:
\be
V^{1/4} / \ex^{1/4} \simeq 7 \times 10^{16}~{\rm GeV}.
\label{fourteen} \ee

In order to determine the range of field values 
that lead to 
inflation, we start by considering a region of space,
the size of a few Hubble lengths, in which the fields are
homogeneous. We determine the evolution of this region by 
integrating eqs. (\ref{three})--(\ref{five}) numerically.
We consider field values $0 \leq \phi,\sigma \leq \mpl$,
and couplings $\lx=g=1$. The choice of couplings has a small effect 
on our conclusions, as long as they are not much smaller than 1. 
Since we are interested in the necessary conditions
for the onset of inflation, we
neglect the term $- m^2 \phi$ in eq. (\ref{four}) during the numerical 
integration. We assume that inflation occurs if the evolution stops
on the $\phi$ axis beyond the instability point. If there is a significant
slope along the $\phi$ axis, the range of field values that lead to inflation
is smaller than our predictions.

In fig. 1 we display an array of initial field values in Planck units
for a theory with $M=10^{-1} \mpl$. 
The dark hexagons denote initial conditions that lead to 
inflation along the positive $\phi$ axis. The trajectories are short,
with the $\phi$ field moving directly to 
its final value, while the $\sigma$ field performs damped oscillations around
zero. 
The white hexagons
denote non-inflationary trajectories that end up at the minimum at
($\phi=0,\sigma=M$). The white triangles
denote trajectories that end up at the minimum at
($\phi=0,\sigma=-M$). 
The white squares at
the beginning of the $\phi$ axis denote non-inflationary trajectories 
that start below the instability point 
($\phi=\sqrt{{\lx}/{g}} M,\sigma=0$) and end up at one of the two minima 
through the growth of fluctuations of the $\phi$ field.
We observe a large continuous range of initial
conditions that lead to the onset of inflation. This range starts 
at the instability point.
  
In fig. 2 we display a similar array of initial field values for
$M=3 \times 10^{-2} \mpl$. The black triangles indicate 
initial values that lead to inflation along the negative 
$\phi$ axis. We observe that the large continuous range of inflationary 
initial values has moved towards the Planck scale. A thin strip remains
along the $\phi$ axis that ends at the instability point.
Trajectories starting in the rest of field space rarely 
lead to the onset of inflation. Both fields oscillate around zero 
before the system relaxes either at one of the minima or 
in the valleys along the $\phi$ axis.  
In fig. 3 we present a magnification of fig. 2 around the origin,
which demonstrates the
size of the various areas that lead to inflation. The largest of these
has a size $\sim 2$--$3 \times 10^{-2} \mpl$, but the typical size is
$\sim 10^{-2} \mpl$. The thin strip around the $\phi$ axis has
a width $\sim 5 \times 10^{-3} \mpl$. Another magnification of 
fig. 2, away from the origin, is presented in fig. 4.
Inflationary strips of width $\sim 10^{-2} \mpl$ are visible. 

Another array of initial field values,
for $M= 10^{-2} \mpl$, is presented in fig. 5.
The situation is similar to that of fig. 2, but the large
inflationary range has moved further towards the Planck scale. 
Figures 6 and 7 are magnifications of fig. 5 that demonstrate the 
size of the various areas. The inflationary strips of fig. 6
have a
width $\sim 1$--$2 \times 10^{-6} \mpl$. Both fields oscillate many times 
before the evolution stops.
The strip along the
$\phi$ axis has a width  $\sim 4 \times 10^{-4} \mpl$. 

The shrinking of the areas of field space that lead to inflation
is due to the decrease of $H$ during the last stages of 
the evolution as $M$ is lowered. This reduces the magnitude
of the friction term in eqs. (\ref{four}) and (\ref{five}). 
Consequently, it takes longer for the system to settle down 
either at the minima or
in the valleys along the $\phi$ axis. Many oscillations
of the fields are required before most of the initial potential energy
is dissipated through expansion. 
This results is increasing sensitivity to the 
initial conditions. A slight change of the initial field values
causes the system to move off an inflationary trajectory and 
end up at one of the minima instead. 
For $M \lta 10^{-2} \mpl$ the largest area that leads to 
inflation, for initial field values not very close to the Planck scale, 
is the strip along the $\phi$ axis. There, the 
evolution is rather simple, with only the $\sigma$ field oscillating and
$\phi$ moving directly to its final value.

We can obtain an estimate of the width of this area if we consider 
eq. (\ref{four}) with $m^2=0$ and substitute for $\sigma^2$ its
average value $\sigma^2_{rms}=\sigma^2/2$.
There are two time scales
characterizing the solutions of this equation.
The first one is related to the friction term and is given by  
$t_H^{-1} = 3H/2$. 
For $\sigma \ll M$ and $\phi/M \ll \sqrt{\lx/g}~M/\sigma$, we
have 
\be
t_H =  \sqrt{\frac{2}{3 \pi \lx}}  \frac{\mpl}{M^2}.
\label{fifteen} \ee
The other time scale is obtained if we neglect the friction term and 
consider the oscillations of the $\phi$ field.
One-fourth of the period gives the typical time for the system to roll
to the origin and away from an inflationary solution. It is given
by 
\be 
t_{osc} = \frac{\pi}{\sqrt{2g}} \frac{1}{\sigma}.
\label{sixteen} \ee
Inflation sets in if $t_{osc} > t_H$, which gives 
\be
\frac{\sigma}{\mpl} < 
\sqrt{\frac{3 \pi^3 \lx}{4g}}\left( \frac{M}{\mpl} \right)^2.
\label{seventeen} \ee
For comparable couplings $\lx$, $g$, 
our assumptions for the derivation of the above bound break 
down when either $M \sim \mpl$ or 
$\phi \gta \sqrt{4/3\pi^3}  \mpl \simeq 0.2 \mpl$.
This approximate analysis is verified by the results in 
figs. 3 and 7. We have also checked eq. (\ref{seventeen}) numerically 
for $M=10^{-3} \mpl$. For this value, the large continuous range of
inflationary initial conditions moves to $\phi$ field values 
above $\mpl$.

The implications for the initial configuration that will lead to the
onset of inflation are severe. For initial field values below $\mpl$
and $M \lta 10^{-3} \mpl$, the
most favourable area of inflationary initial conditions is
around the $\phi$ axis. 
Throughout a region of space the size of the order of the  
Hubble length (which initially is 
$\sim \mpl^{-1}$), the spatial variation of the $\sigma$ 
field in Planck units
cannot exceed the r.h.s. of eq. (\ref{seventeen}). 
If this condition is not satisfied, the fields in
different parts of the original
homogeneous space region will evolve towards very different values.
In one part they may end up in the valleys along the $\phi$ axis,
while in another they may settle at the minima of the potential. 
Before inflation, the 
size of space regions shrinks compared to the
Hubble distance (for a 
quantitative discussion, see below). 
As a result, large inhomogeneities are expected at scales smaller than
$\sim H^{-1}$ when the evolution of the fields finally
slows down. These will prevent the onset of inflation. 
We conclude that,
for realistic inflationary models (that require $M \lta 10^{-3} \mpl$),
a significant fine tuning is required for the initial configuration 
that will lead to inflation.

The argument that we have presented is not rigorous. 
In principle, one should
integrate the field equations for a configuration with a small
spatial dependence and observe the growth of the perturbation \cite{gold}. 
However, we believe that the approximation of
considering the initial space region as being composed of a few
smaller regions with slightly different constant field values captures
the essence of the problem. For this reason, we expect that 
a more rigorous treatment will not invalidate our 
qualitative conclusions. 

Up till now we have not considered the kinetic energy density 
of the fields. We have assumed that the fields evolve
from a homogeneous value with zero time derivatives. 
This is not a good approximation when the potential
is much smaller than $\mpl^4$. A more reasonable approximation
is to fix the total initial energy density at $\mpl^4$, and
consider a kinetic term equal to the difference between
the total energy density and the potential. However, this does not
fix the direction of the initial time derivative of the fields. 
If this direction is uniform throughout the initial space 
region, we do not expect a qualitative change in the behaviour we have
observed previously \cite{gold}. 
In fig. 8 we display an array of initial field values in Planck units
for a theory with $M=10^{-2} \mpl$. The initial energy density has
been fixed at $E=10^{-4} \mpl^4$. 
The initial time derivatives of
the fields have been chosen equal and positive. We observe 
a strong similarity between figs. 6 and 8. This is explained 
by the fact that the initial time derivative quicly causes the 
displacement of the fields
from the $\phi$ axis into an area where $\phi,\sigma \sim 0.1 \mpl$.
There, the pattern of fig. 6
is expected. The larger inflationary strip of fig. 7 has disappeared.
If the initial energy density is taken $E \sim \mpl^4$, 
the fields are displaced to regions $\phi,\sigma \sim \mpl$, where
the inflationary strips are narrower than the ones depicted in fig. 8. 
We conclude that the presence of an initial kinetic term imposes much
stronger constraints on the initial conditions than that
of eq. (\ref{seventeen}). There is the exception of
an initial time derivative along the $\phi$ axis. However, this possibility
leads to field values larger than $\mpl$. 
The variation of 
the direction of the time derivative within the initial space region 
results in more irregular patterns than that of fig. 8, and 
strengthens the constraints further. 

We have also followed the evolution of the scale 
factor $R$ relative to the Hubble parameter, 
starting from an initial value 
$R_0 \sim H^{-1}_0$. For the initial conditions of fig. 8 
that lead to inflation, we find that at the onset of
inflation $R^{-1}$ is typically smaller than $H^{-1}$ by a factor 
of order 10.
For initial energy densities $E \sim \mpl^4$, this factor becomes 
of order 100.
This implies that the initial homogeneous region should extend far 
beyond a few initial Hubble lengths for this region to inflate.
This factor becomes of order 1 if the initial time derivative
of the fields is small and the fields start near the $\phi$ axis and
move quickly on it.

We have based our discussion on the integration of
the classical equations of motion.
The presence of quantum (or thermal) fluctuations of the fields around
their average values can be treated as a stochastic noise.
We expect that this contributes an additional destabilizing factor,
as it can easily knock the fields off an inflationary 
trajectory and enhance the irregularity of patterns in figs. 1--8.

We conclude that hybrid inflation with the potential of 
eq. (\ref{one}) is not ``natural''. An extremely homogeneous
initial field configuration must be assumed for initial field values
below $\mpl.$ For realistic models with  $M \lta 10^{-3}\mpl$,
the $\sigma$ field must be homogeneous with 
at least $\sim 10^{-6}$--$10^{-5}$ accuracy. 
This bound may be strengthened by several orders of 
magnitude if the initial fields have a significant 
time derivative. These constraints do not apply if the initial values
of the fields are larger than $\mpl$. One could imagine that inflation 
starts in this region and continues as the system rolls
down the valley on the $\phi$ axis.  
It is not possible, however, to 
construct reliable field-theoretical models in the field range above $\mpl$.

Our results have direct application to 
the simplest supersymmetric model that supports hybrid inflation.
It is described by the 
superpotential \cite{cop,shafi}
\be
W = \Phi \left( -\mu^2 + \lx \bar{\Psi} \Psi \right).
\label{twoone} \ee
Here $\Phi$, $\Psi$ and $\bar{\Psi}$ 
are chiral superfields, for which we 
assume canonical kinetic terms. 
The above superpotential is the only renormalizable
one consistent with a continuous $U(1)$ $R$-symmetry 
under which $W \rightarrow e^{i\theta} W$, $\Phi \rightarrow e^{i\theta} \Phi$,
$ \bar{\Psi} \Psi \rightarrow \bar{\Psi} \Psi$. 
The superfields $\Psi$ and $\bar{\Psi}$ 
transform under an internal gauge symmetry,
while the  $\Phi$ superfield is a 
gauge singlet. 
The gauge symmetry is
spontaneously broken
at the scale $\mu/\sqrt{\lx}$. The scale of inflation is
set by $\mu$ and is constrained by the COBE observations 
to be two or three orders of magnitude below $\mpl$.  

The potential of the scalar components of the superfields (for which
we use the same notation) is given by 
\be
V = \left| -\mu^2 + \lx \bar{\Psi} \Psi \right|^2 + 
\lx^2 \left| \Phi \right|^2
\left( \left| \Psi \right|^2 + \left| \bar{\Psi} \right|^2 \right)
+ D{\rm-terms}.
\label{twotwo} \ee
Vanishing of the $D$-terms is achieved along the $D$-flat directions 
where $\left| \Psi \right| = \left| \bar{\Psi} \right|$. 
If the fields are assumed real, the potential along the $D$-flat directions
is completely analogous to that of eq. (\ref{one}) but without a mass term
for $\Phi$. 
The slope along the 
inflationary trajectory can be generated either through the addition of
a soft supersymmetry-breaking mass for $\Phi$ \cite{cop} or by logarithmic 
radiative corrections \cite{shafi}. Our conclusions about the 
fine-tuning of the initial configuration that will lead to the
onset of inflation are independent of the term providing this
slope. The figures 1--8 were generated for zero slope, and the 
constraints that we derived are strengthened if a significant
slope is present.   
Including the imaginary parts of the scalar fields (some of which 
can be eliminated through appropriate gauge and $R$-transformations 
along the $D$-flat directions) does not alter the qualitative picture. 

It must be pointed out, however, that our conclusions 
do not automatically apply to models with 
superpotentials different from that 
of eq. (\ref{twoone}). 
Models such as the smooth hybrid inflation of
ref. \cite{giwrgos} may display less sensitivity on the initial
conditions and their separate investigation is necessary.

\vspace{0.5cm}
\noindent
{\bf Acknowledgements}: 
We would like to thank J. Garcia-Bellido, 
G. Lazarides and C. Panagiotakopoulos for many useful discussions.

\newpage

\pagestyle{empty}

\begin{figure}
\psfig{figure=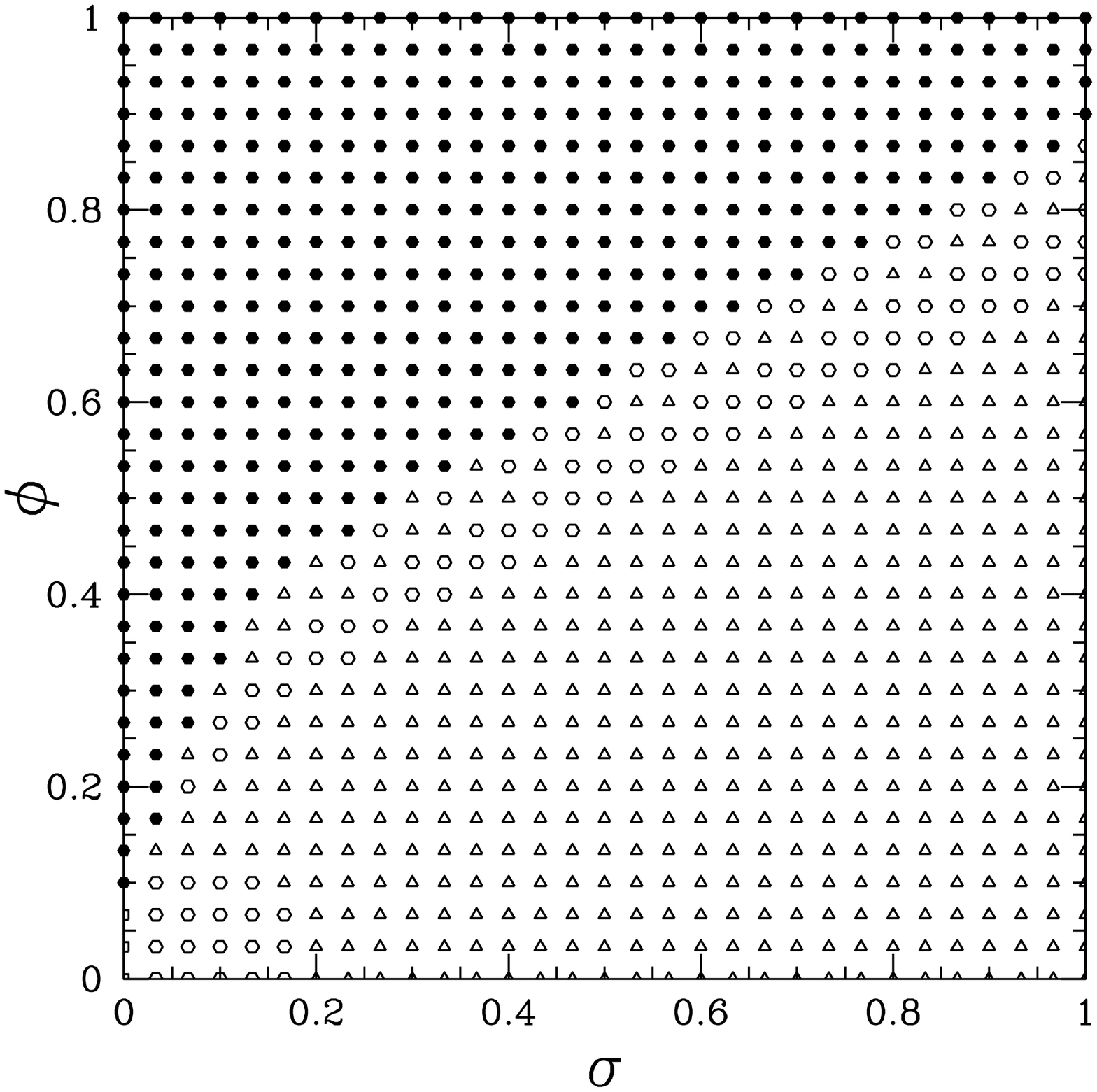,height=16.0cm}
\vspace*{2cm}
Fig. 1: $M=10^{-1}\mpl$, $\lx=1$, $g=1$. 
\end{figure}

\begin{figure}
\psfig{figure=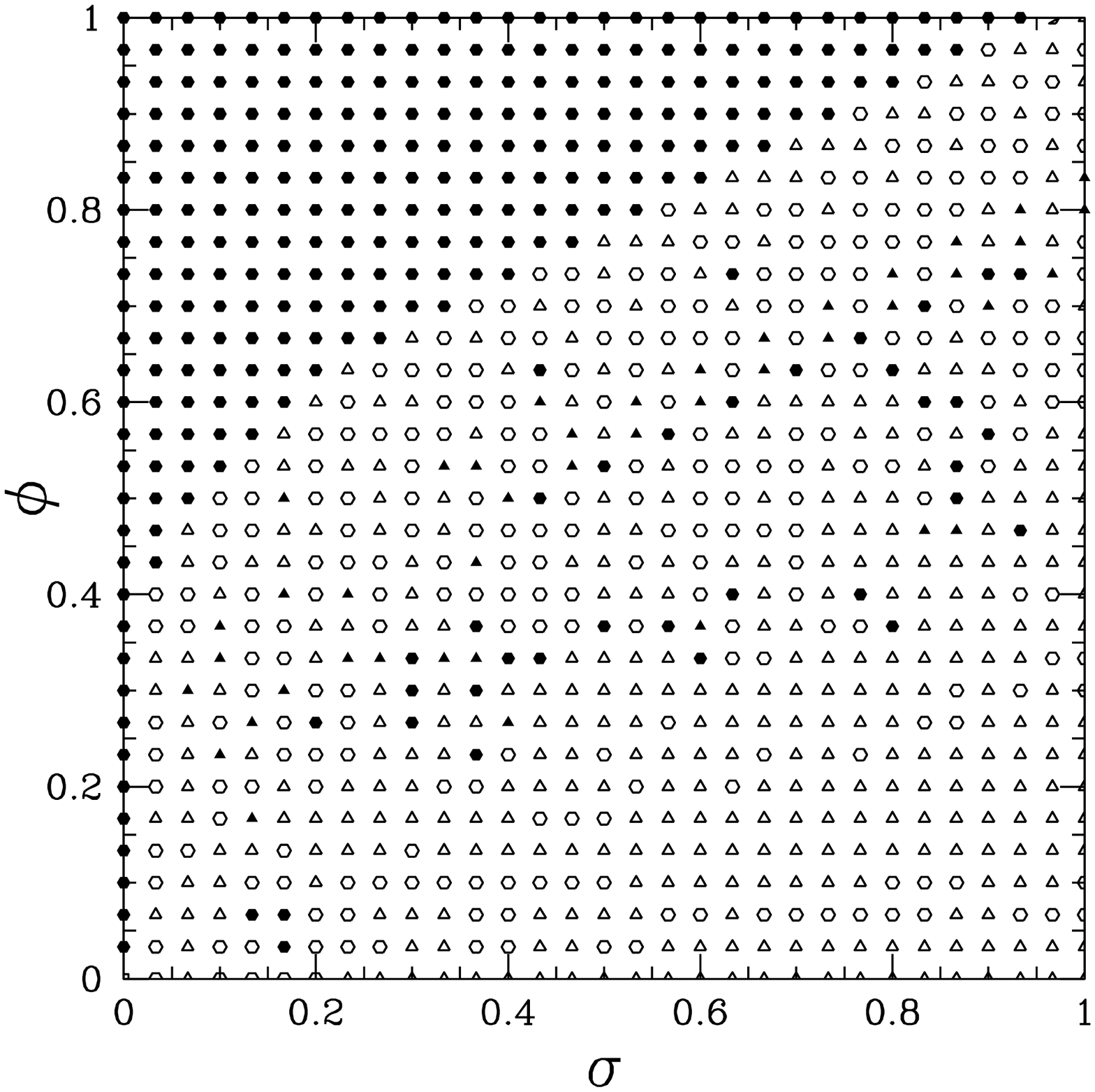,height=16.0cm}
\vspace*{2cm}
Fig. 2: $M=3 \times 10^{-2}\mpl$, $\lx=1$, $g=1$. 
\end{figure}

\begin{figure}
\psfig{figure=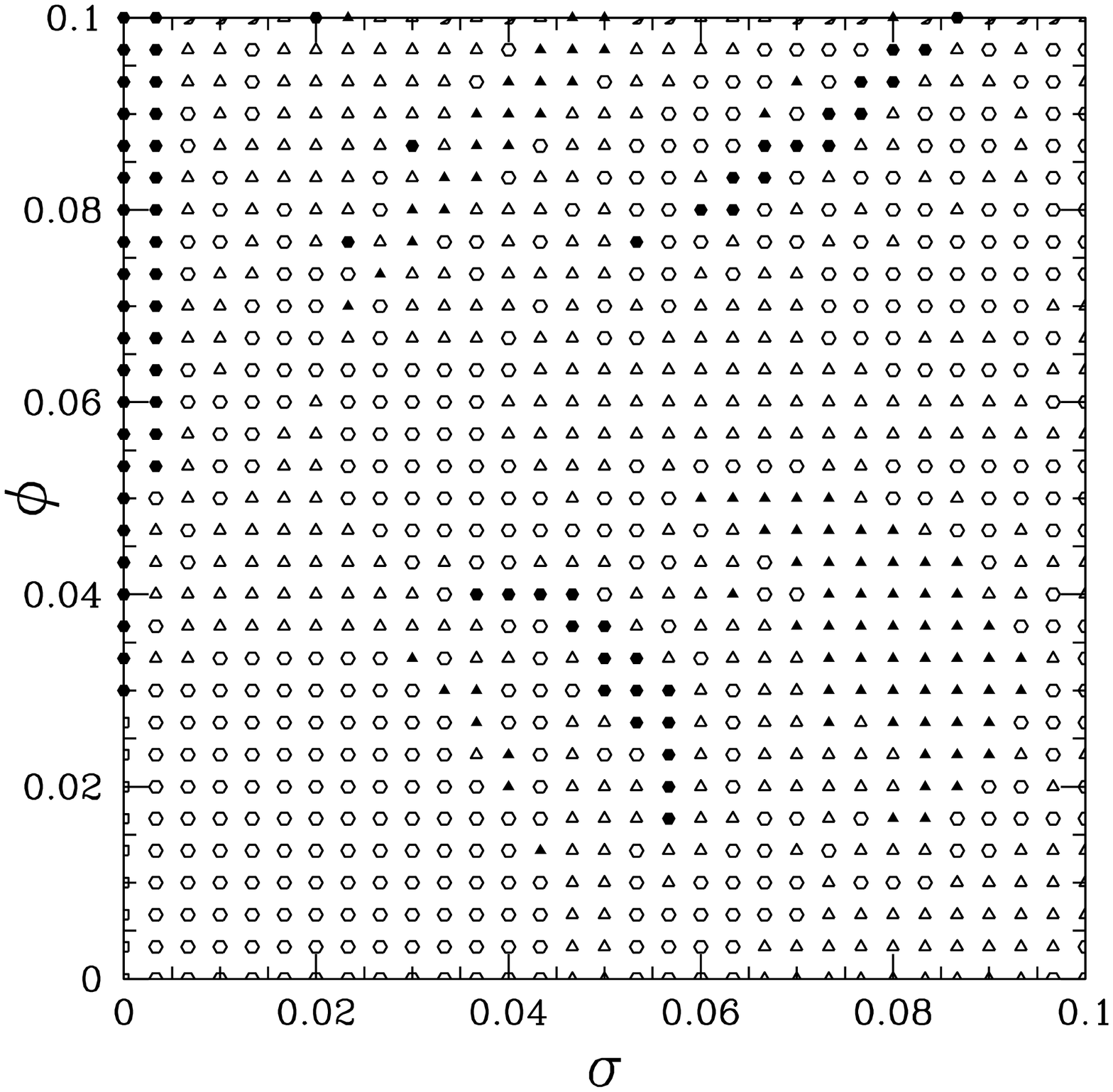,height=16.0cm}
\vspace*{2cm}
Fig. 3: Magnification of fig. 2.
\end{figure}

\begin{figure}
\psfig{figure=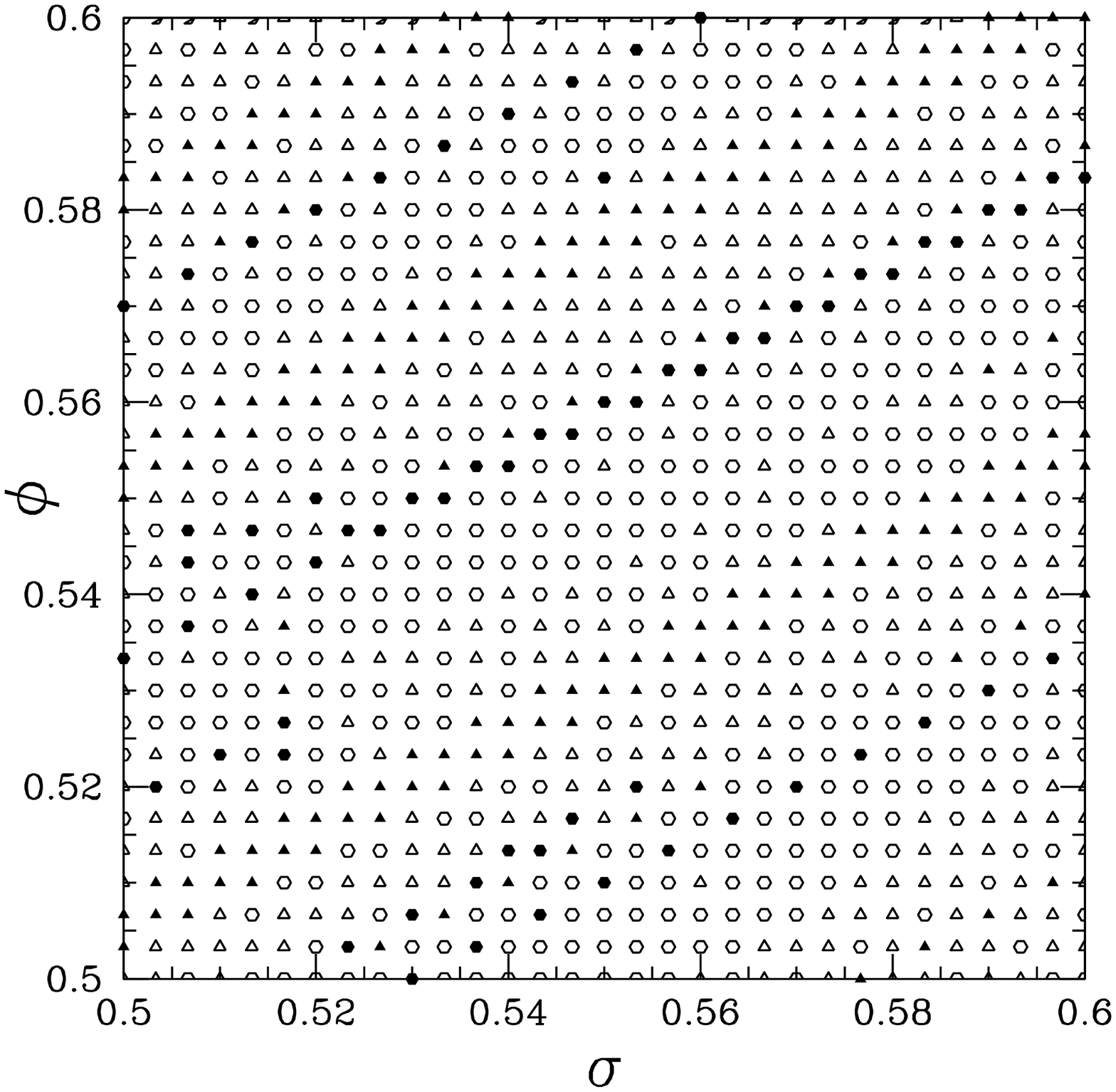,height=16.0cm}
\vspace*{2cm}
Fig. 4: Magnification of fig. 2.
\end{figure}

\begin{figure}
\psfig{figure=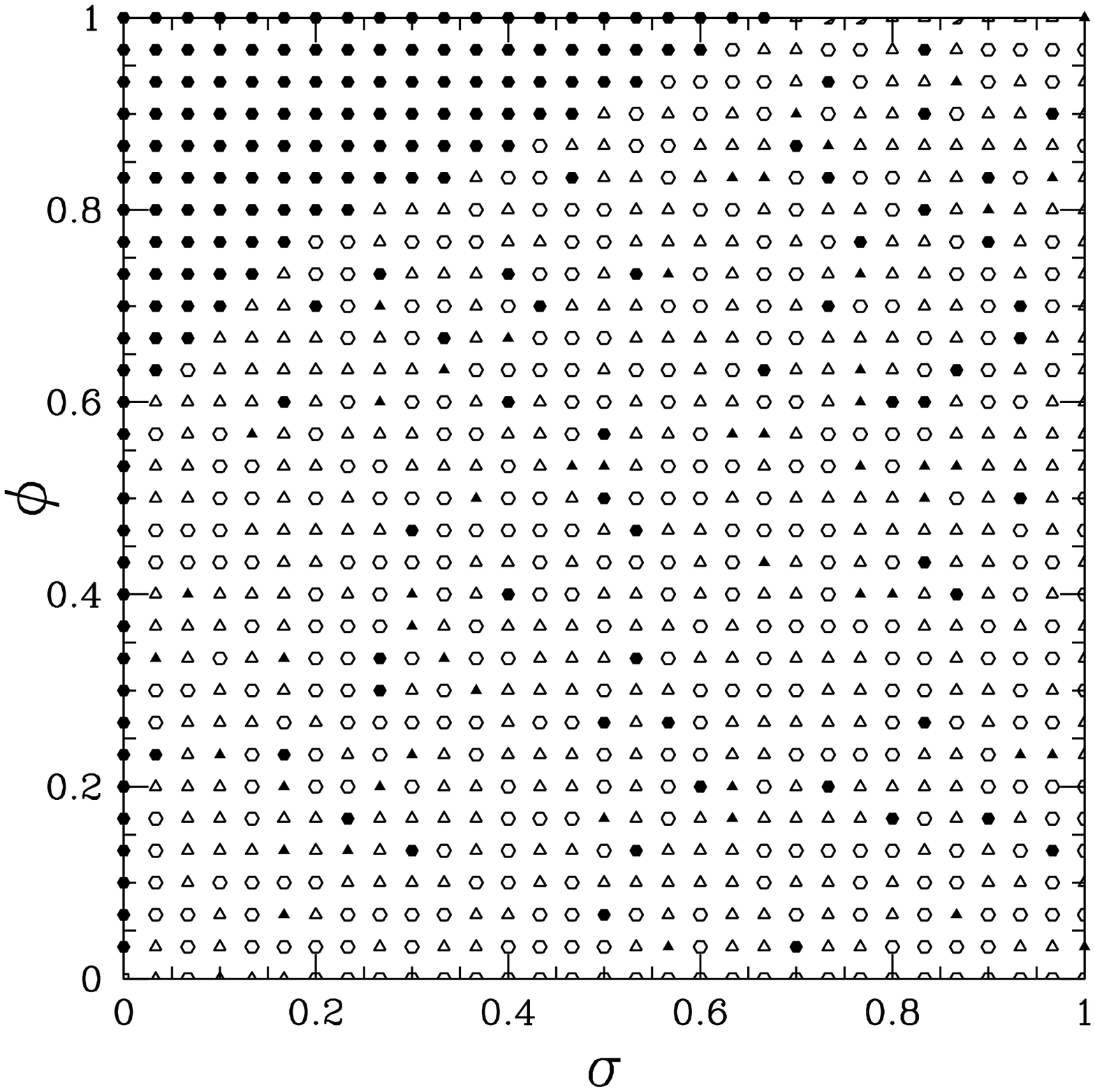,height=16.0cm}
\vspace*{2cm}
Fig. 5: $M= 10^{-2}\mpl$, $\lx=1$, $g=1$. 
\end{figure}

\begin{figure}
\psfig{figure=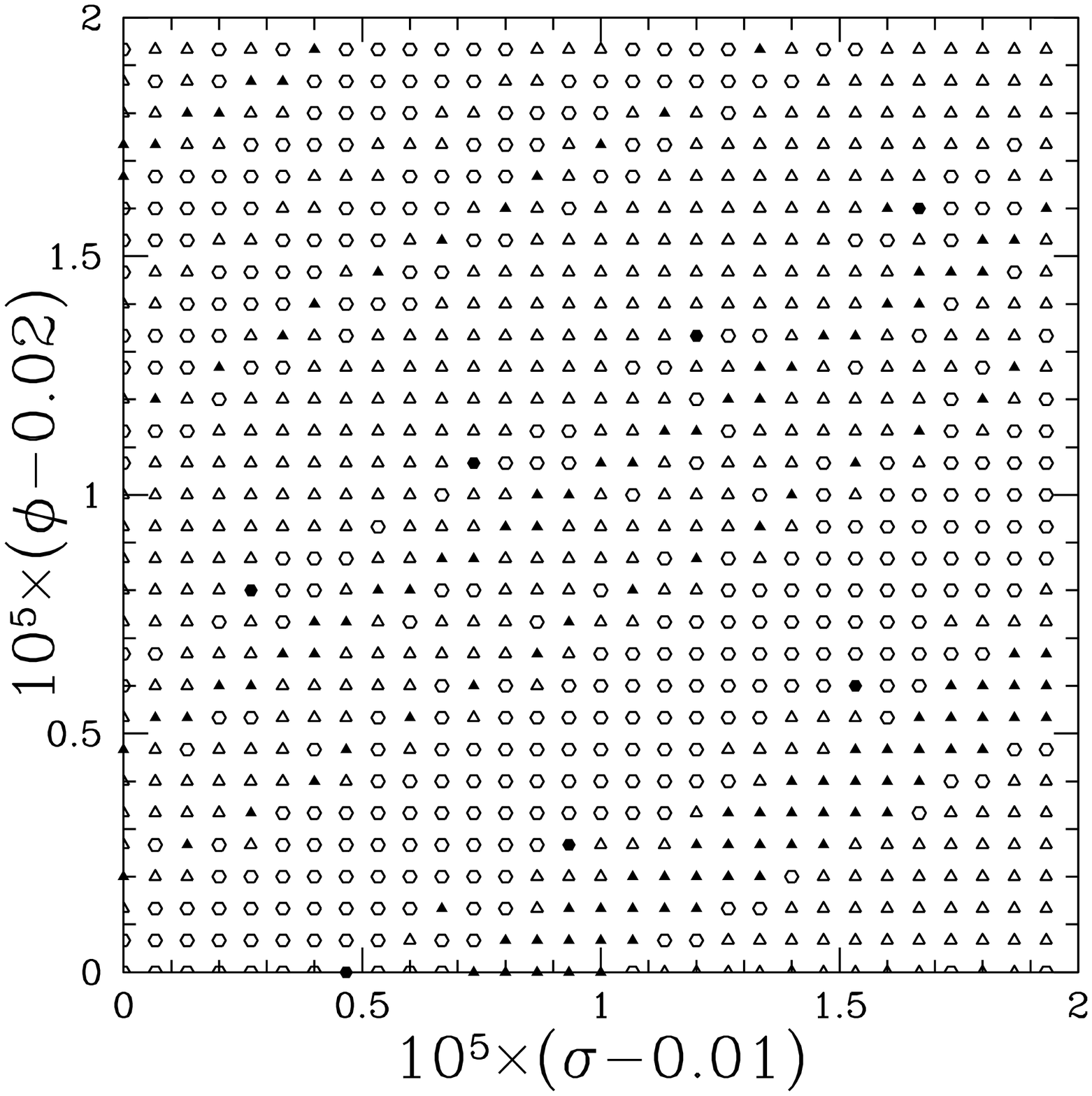,height=16.0cm}
\vspace*{2cm}
Fig. 6: Magnification of fig. 5.
\end{figure}

\begin{figure}
\psfig{figure=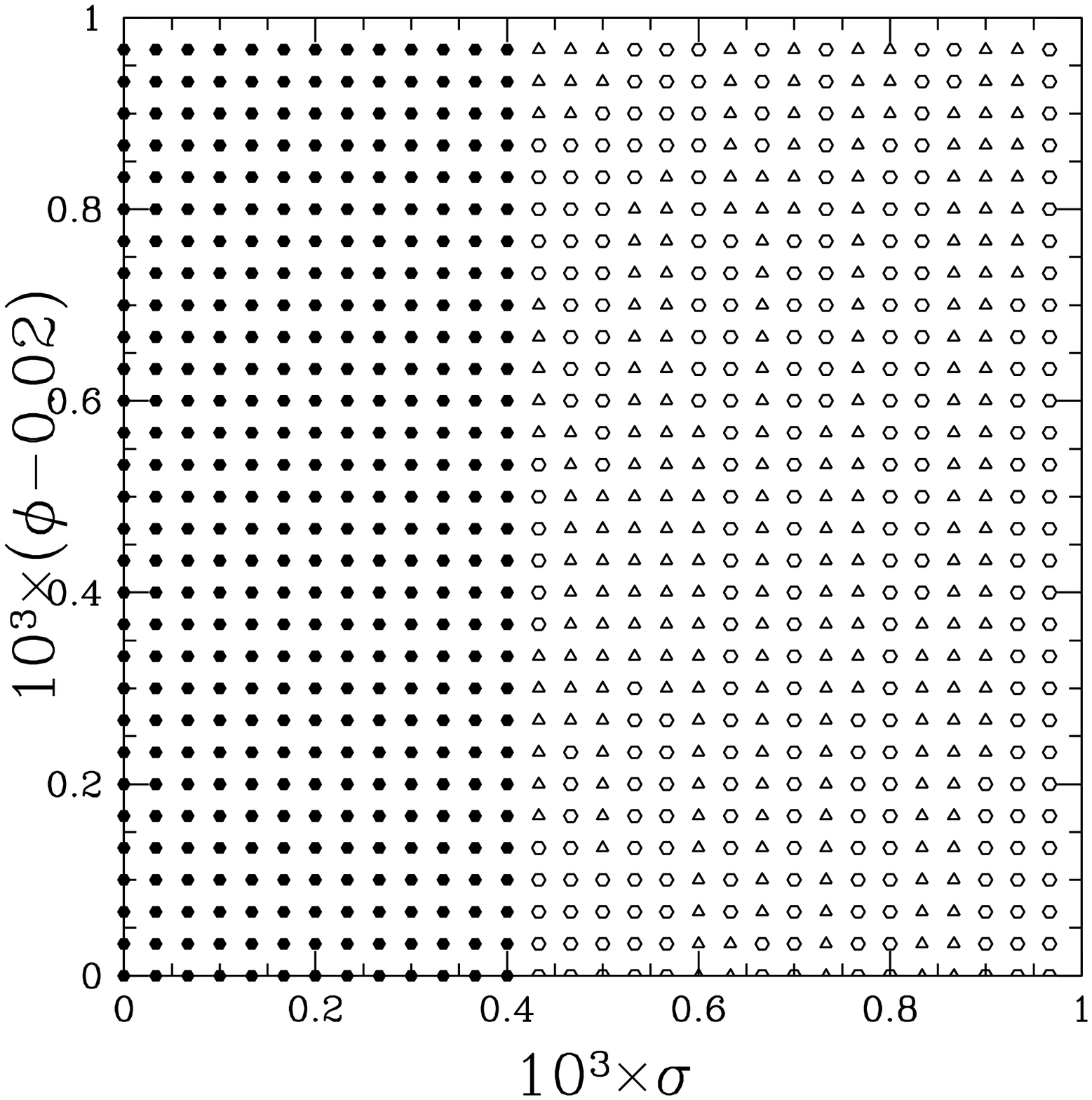,height=16.0cm}
\vspace*{2cm}
Fig. 7: Magnification of fig. 5.
\end{figure}

\begin{figure}
\psfig{figure=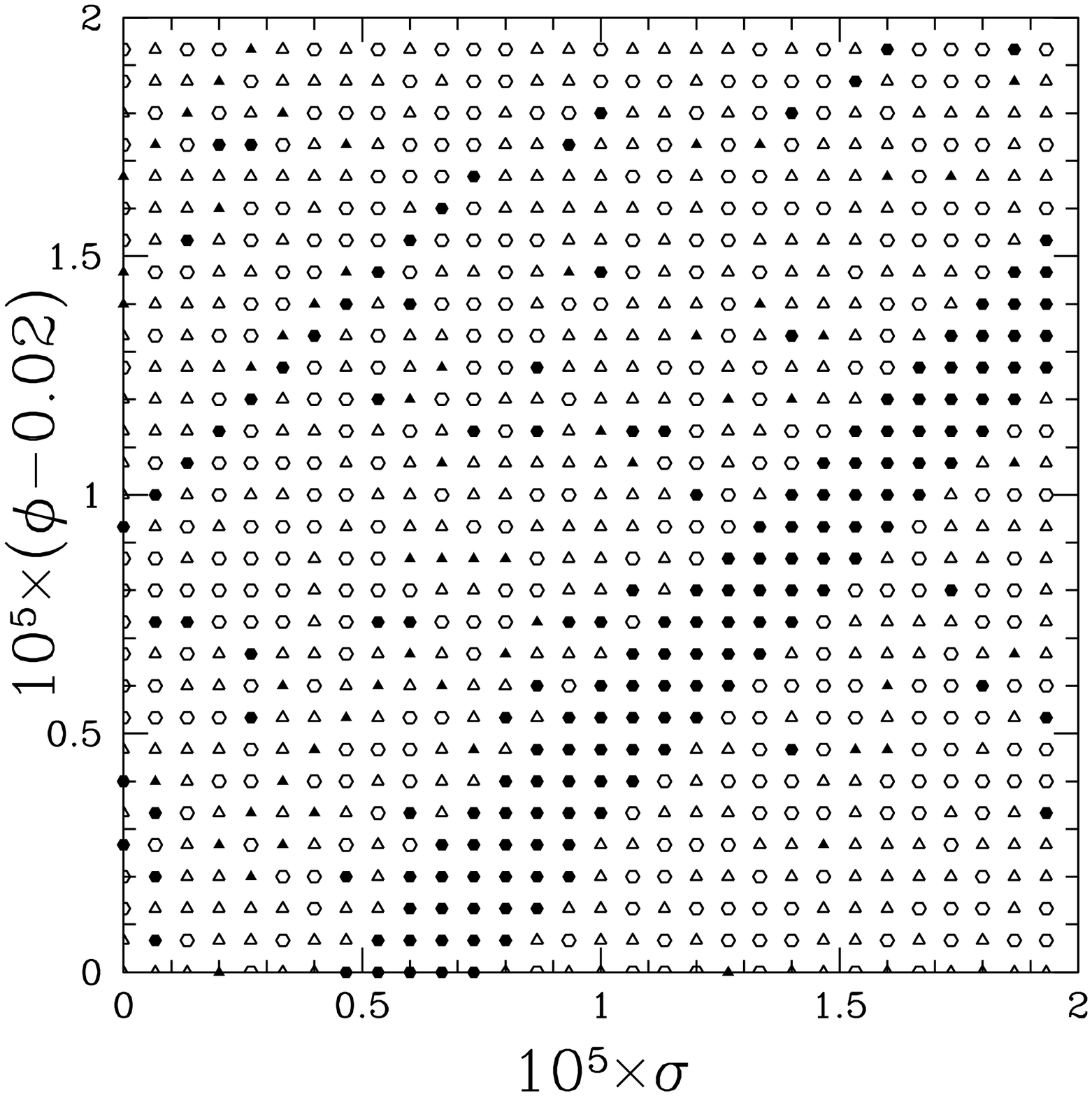,height=16.0cm}
\vspace*{2cm}
Fig. 8: $M= 10^{-2}\mpl$, $\lx=1$, $g=1$. Initial total energy
density $E = 10^{-4} \mpl^4$ and $\dot{\phi}=\dot{\sigma}>0$.
\end{figure}


\newpage




\begin{thebibliography}{99}

\bibitem{hybrid}
A.D. Linde, Phys. Lett. B {\bf 259}, 38 (1991);
Phys. Rev. D {\bf 49}, 748 (1994). 

\bibitem{cop}
E.J. Copeland, A.R. Liddle, D.H. Lyth, E.D. Stewart and D. Wands,
Phys. Rev. D {\bf 49}, 6410 (1994).

\bibitem{shafi}
G. Dvali, Q. Shafi and R. Schaefer, Phys. Rev. Lett. {\bf 73}, 1886
(1994).
 
\bibitem{stewart} 
E.D. Stewart, Phys. Rev. D {\bf 51}, 6847 (1995).

\bibitem{giwrgos}
G. Lazarides and C. Panagiotakopoulos,
Phys. Rev. D {\bf 52}, 559 (1995).

\bibitem{costas}
C. Panagiotakopoulos, Phys. Rev. D {\bf 55}, 7335 (1997);
Phys. Lett. B {\bf 402}, 257 (1997).

\bibitem{dterm}
P. Bin\'etruy and G. Dvali, Phys. Lett. B {\bf 388}, 241 (1996);
E. Halyo, Phys. Lett. B {\bf 387}, 43 (1996).

\bibitem{riotto}
A.D. Linde and A. Riotto, Phys. Rev. D {\bf 56}, 1841 (1997).

\bibitem{gia}
S. Dimopoulos, G. Dvali and R. Rattazzi,
Phys. Lett. B {\bf 410}, 119 (1997).

\bibitem{first1}
G. Lazarides, C. Panagiotakopoulos and N.D. Vlachos, 
Phys. Rev. D {\bf 54}, 1369 (1996).

\bibitem{first2}
G. Lazarides and N.D. Vlachos, Phys. Rev. D {\bf 56},
4562 (1997). 

\bibitem{chaotic}
A.D. Linde, Particle physics and inflationary cosmology 
(Harwood Academic Publishers, Chur, 1990), and references therein.

\bibitem{report}
A.R. Liddle and D.H. Lyth, Phys. Rep. {\bf 231}, 1 (1993). 

\bibitem{denpert}
S.W. Hawking, Phys. Lett. B {\bf 115}, 295 (1982);
A.A. Starobinsky, Phys. Lett. B {\bf 117}, 175 (1982);
A.H. Guth and S.-Y. Pi, Phys. Rev. Lett. {\bf 49}, 1110 (1982);
D.H. Lyth, Phys. Lett. B {\bf 147}, 403 (1984); {\it ibid.}
{\bf 150}, 465 (1985); Phys. Rev. D {\bf 31}, 1792 (1985).

\bibitem{lyth}
D.H. Lyth, preprint LANCS-TH/9614, hep-ph/9609431.

\bibitem{gold}
D.S. Goldwirth and T. Piran,
Phys. Rep. {\bf 214}, 223 (1992).


\end{thebibliography}
\end{document}